\documentclass[usenatbib]{aa}

\usepackage{comment}
\usepackage{placeins}
\usepackage{graphicx}
\usepackage{amsmath,bm}
\usepackage{wasysym}
\usepackage{xspace}
\usepackage{xcolor}
\usepackage{soul}
\usepackage{natbib}
\usepackage{txfonts}
\usepackage{adjustbox}
\usepackage[version=4]{mhchem}
\bibpunct{(}{)}{;}{a}{}{,}

\usepackage{txfonts,textcomp}

\usepackage[colorlinks=true,allcolors=blue]{hyperref}
\usepackage{etoolbox}

\makeatletter
  \patchcmd{\NAT@citex}
    {\@citea\NAT@hyper@{
      \NAT@nmfmt{\NAT@nm}
      \hyper@natlinkbreak{\NAT@aysep\NAT@spacechar}{\@citeb\@extra@b@citeb}
      \NAT@date}}
    {\@citea\NAT@nmfmt{\NAT@nm}
    \NAT@aysep\NAT@spacechar\NAT@hyper@{\NAT@date}}{}{}

  \patchcmd{\NAT@citex}
    {\@citea\NAT@hyper@{
      \NAT@nmfmt{\NAT@nm}
      \hyper@natlinkbreak{\NAT@spacechar\NAT@@open\if*#1*\else#1\NAT@spacechar\fi}
        {\@citeb\@extra@b@citeb}
      \NAT@date}}
    {\@citea\NAT@nmfmt{\NAT@nm}
    \NAT@spacechar\NAT@@open\if*#1*\else#1\NAT@spacechar\fi\NAT@hyper@{\NAT@date}}
    {}{}
\makeatother

\usepackage{soul}
\usepackage{adjustbox}

\newcommand{\Msun}{\,{\rm M_{\odot}}}

\newcommand{\epsff}{\epsilon_{\mathrm{ff}}}

\newcommand\orcid[1]{\protect\href{http://orcid.org/#1}{\includegraphics[height=12pt]{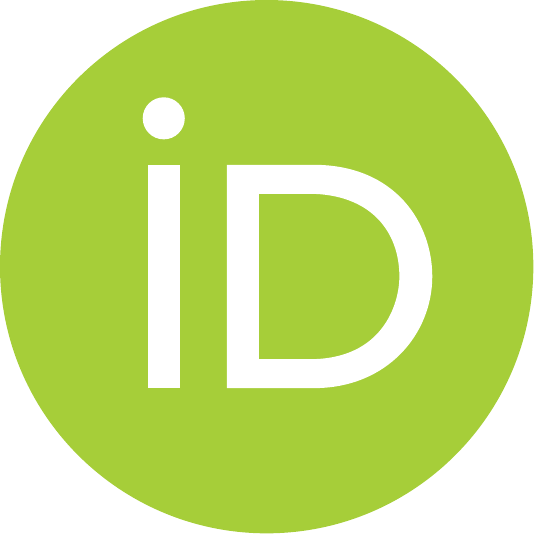}}}

\begin{document}

   \title{The life cycle of giant molecular clouds in simulated Milky Way-mass galaxies}
   \authorrunning{Y. Ni et al.}

   \author{Yang Ni \inst{1,3} \orcid{0000-0003-0794-1949}
       \and
           Hui Li \inst{2} \orcid{0000-0002-1253-2763}
       \and
           Mark Vogelsberger \inst{3} \orcid{0000-0001-8593-7692}
        \and
           Laura V. Sales \inst{4} \orcid{0000-0002-3790-720X}
        \and
           Federico Marinacci \inst{5,6} \orcid{0000-0003-3816-7028}
        \and
           Paul Torrey \inst{7} \orcid{0000-0002-5653-0786}
    }

   \institute{Institute for Advanced Study, Tsinghua University, Beijing 100084, People’s Republic of China
         \and
             Department of Astronomy, Tsinghua University, Beijing 100084, People’s Republic of China\\
             \email{\href{mailto:hliastro@tsinghua.edu.cn}{hliastro@tsinghua.edu.cn}}
        \and
            Department of Physics and Kavli Institute for Astrophysics and Space Research, Massachusetts Institute of Technology, Cambridge, MA 02139, USA
        \and
             Department of Physics and Astronomy, University of California, Riverside, CA 92521, USA
        \and
             Department of Physics and Astronomy "Augusto Righi", University of Bologna, Via P. Gobetti 93/2, I-40129 Bologna, Italy
         \and
             INAF, Astrophysics and Space Science Observatory Bologna, Via P. Gobetti 93/3, I-40129 Bologna, Italy
         \and
             Department of Astronomy, University of Virginia, Charlottesville, VA 22904, USA
             }
    \date{Received Month DD, Year; accepted Month DD, Year}

   \abstract
   {Giant molecular clouds (GMCs) are the primary sites of star formation in galaxies. Their evolution, driven by the interplay of gravitational collapse, stellar feedback, and galactic dynamics, is key to understanding local star formation on GMC scales. However, tracking the full life cycle of GMCs across diverse galactic environments remains challenging and requires high-resolution hydrodynamical simulations and robust post-processing analysis.}
   {We aim to trace the complete life cycle of individual GMCs in high-resolution Milky Way–mass galaxy simulations to determine how different stellar feedback mechanisms and galactic-scale processes govern cloud lifetimes, mass evolution, and local star formation efficiency (SFE).}
   {We identified GMCs in simulated galaxies and tracked their evolution using cloud evolution trees. Via cloud evolution trees, we quantified the lifetimes and SFE of GMCs. We further applied our diagnostics to a suite of simulations with varying star formation and stellar feedback subgrid models and explored their impact together with galactic environments to the GMC life cycles.}
   {
   Our analysis reveals that GMCs undergo dynamic evolution, characterized by continuous gas accretion, gravitational collapse, and star formation, followed by disruption due to stellar feedback. The accretion process sustains the gas content throughout most of the GMC life cycles, resulting in a positive correlation between GMC lifetimes and their maximum masses. The GMC lifetimes range from a few to several tens of million years, with two distinct dynamical modes: (1) GMCs near the galactic center experience strong tidal disturbances, prolonging their lifetimes when they remain marginally unbound; (2) those in the outer regions are less affected by tides, remain gravitationally bound, and evolve more rapidly. In all model variations, we observe that GMC-scale SFE correlates with the baryonic surface density of GMCs, consistent with previous studies of isolated GMCs. Additionally, we emphasize the critical role of galactic shear in regulating GMC-scale star formation and refine the correlation between local SFE and surface density by including its effects. These findings demonstrate how stellar feedback and galactic-scale dynamics jointly shape GMC-scale star formation in realistic galactic environments.
   }
   {}   


   \maketitle

\section{Introduction}\label{sec:intro}

Star formation is known to be inefficient on a galactic scale, with a global molecular gas depletion time around $2-10\,\mathrm{Gyr}$, roughly one order of magnitude longer than the dynamical timescale of galactic disks and two orders of magnitude longer than the local free-fall time of giant molecular clouds (GMCs) \citep{Kennicutt_1998, Bigiel_2011, Kennicutt_2012AR,Sun_2023}. The long gas depletion timescale as well as low global star formation efficiency (SFE) in galaxies can be explained either by a uniformly low star formation rate across all individual GMCs \citep{Krumholz_2005ApJ, Federrath_2012ApJ, Pokhrel_2021ApJ} or by a highly variable SFE during the evolution of GMCs with diverse GMC properties and galactic environments \citep{Lee_2016ApJ, Kruijssen_2018MNRAS, Chevance_2023}. To distinguish between the above two scenarios, it is critical to investigate the detailed life cycles of GMCs and the local star formation process observationally and theoretically. 

In observations of GMCs in the Local Group, the star formation efficiency per free-fall time—measured through tracers such as H$\alpha$ and FUV—exhibits significant scatter, often exceeding 0.3 dex \citep{Murray_2011ApJ, Lee_2016ApJ, Vutisalchavakul_2016ApJ, Ochsendorf_2017ApJ, Sun_2023}. This observed variability presents a challenge to the theory of a universally low, constant SFE. Moreover, the uniformly low SFE scenario implies long-lived GMCs with a lifetime of at least $\sim100$ Myr. However, numerous observational studies estimate that the GMC lifetimes are significantly shorter, typically in the range of $\sim 5$ Myr to $\sim50$ Myr \citep{Murray_2011ApJ, Miura_2012ApJ, Kruijssen_2014MNRAS, Kruijssen_2018MNRAS, Kruijssen_2019Natur, Chevance_2020MNRAS, Sun_2022AJ, Ward_2022MNRAS}. These lifetimes align with the typical cloud-scale free-fall time and turbulence crossing time, indicating that GMCs are unlikely to sustain low SFE over extended periods. Together, the observed scatter in SFE per free-fall time and the relatively short GMC lifetimes may suggest that the evolution of GMCs is driven by diverse, environmentally dependent dynamical processes, resulting in locally varying SFE across GMCs. Nevertheless, current observational techniques infer local SFE and GMC lifetimes through indirect approaches. Observations can estimate the instantaneous star formation rate using different tracers at present, but this rate is only a ``snapshot'' within the whole GMC life cycle. Additionally, the inferred stellar and gas masses are derived from different tracers, each with its own uncertainties, biases, and limitations.

On the theoretical side, analytical models that consider the combined effects of gravitational collapse, cloud interaction, epicyclic motions, galactic shear, and large-scale gas streaming motions support the GMC lifetimes of several tens of million years \citep{Dobbs_2015MNRAS, Jeffreson_2018MNRAS}. Numerical simulations of isolated GMCs also predict the lifetimes around a few free-fall times and reproduce the observed scatter in GMC-scale SFE, accounting for differences in GMC properties \citep{Grudic_2018MNRAS, Li_2019MNRAS}. However, both analytical models and idealized GMC simulations have notable limitations. For example, analytically predicted GMC lifetimes can be highly sensitive to assumptions about various evolutionary mechanisms, such as gravitational collapse, cloud-cloud collisions, or feedback-driven dispersal. These models often rely on simplified or isolated treatments of each mechanism, potentially overlooking complex interactions and dependencies between them. Furthermore, while numerical simulations capture several key physical processes within isolated GMCs, these studies \citep[e.g.,][]{Murray_2010ApJ, Dale_2013MNRAS, Skinner_2015ApJ, Howard_2016MNRAS, Kim_2017ApJ, Grudic_2018MNRAS, Li_2019MNRAS, Grudic_2021MNRAS} ignore the critical effects on the galactic scale, such as shear from the larger galactic environment and continuous gas accretion from the surrounding medium. These environmental factors can significantly influence the stability, collapse timescale, and mass evolution of GMCs, highlighting the importance of including galactic context for more realistic predictions of GMC lifetimes and SFE variability. 

Over the past few decades, due to the advance in state-of-the-art galaxy formation simulations, especially the implementation of star formation \citep[e.g.,][]{Cen_1992ApJ, Springel_2003MNRAS, Krumholz_2011ApJ, Padoan_2011ApJ, Semenov_2016ApJ, Li_2017ApJ,Dubois_2021A&A, Gensior_2020MNRAS, Valentini_2023MNRAS, Girma_2024MNRAS, Ragone-Figueroa_2024A&A} and stellar feedback \citep[e.g.,][]{Stinson_2006MNRAS, Vogelsberger_2013MNRAS, Ceverino_2014MNRAS, Hopkins_2014MNRAS, Smith_2018MNRAS, Lupi_2019MNRAS, Marinacci_2019MNRAS, Keller_2020MNRAS} subgrid models, galaxy formation simulations start to be able to capture the structures of ISM and treat individual star-forming regions properly \citep{Wang_2015MNRAS, Hopkins_2018MNRAS, Li_2020MNRAS, Li_2022MNRAS, ReinaCampos_2022MNRAS, Nobels_2023arXiv,Hopkins_2023MNRAS,Wibking_2023MNRAS, Zhao_2024ApJ}. Some of these simulations include explicit stellar feedback from multiple channels—such as radiation, stellar winds, and supernovae—which, though not directly resolved, are crucial for modeling GMC properties and their evolution. As a result, these modern simulations now provide new laboratories to track the life cycles of GMCs \citep[e.g.,][]{Benincasa_2020MNRAS, Jeffreson_2021MNRAS, Khullar_2024ApJ} and investigate local SFE in realistic galactic environments.

In this paper, we identified GMCs and tracked the temporal evolution during their life cycles in a set of simulations of isolated Milky Way-like galaxies introduced in \citet[][hereafter L20]{Li_2020MNRAS}. This suite of simulations was performed with the moving-mesh code \textsc{arepo} \citep{Springel_2010MNRAS,2016MNRAS.455.1134P,2020ApJS..248...32W:Weinberger}, using the Stars and MULtiphase Gas in GaLaxiEs (SMUGGLE) model \citep[][hereafter M19]{Marinacci_2019MNRAS} to simulate key ISM processes that are crucial to the life cycles of GMCs. We investigated the lifetime distribution and SFE of the tracked GMC along time series, using the GMC identification in the three-dimensional space and establishing the novel tree network to follow the star formation within GMCs. The paper is organized as follows. In Section~\ref{sec:methods}, we briefly introduce the physical ingredients in the simulations, describe the method to identify and track GMCs, demonstrate the construction of the cloud evolution tree, and illustrate the detection of star formation along the established tree network. In Section~\ref{sec:results}, we examine the properties of the GMCs, calculate their lifetime distribution, categorize the life cycle of the GMCs in different modes, and analyze the star formation process for different model variations. In Section~\ref{sec:discussion}, we compare our results with previous work and derive an analytical model for the SFE including both galactic scale effect and stellar feedback. Finally, in Section~\ref{sec:summary}, we summarize the key results of this work.

\section{Methods}\label{sec:methods}

\subsection{Simulation details and model variations}\label{sec:sim}

The simulations analyzed in this work were introduced in L20 and were performed with the moving-mesh finite-volume hydrodynamic code \textsc{arepo} \citep{Springel_2010MNRAS,2016MNRAS.455.1134P,2020ApJS..248...32W:Weinberger}. The simulations adopted the galaxy formation physics implemented in the SMUGGLE framework, a comprehensive galaxy formation model that includes radiative cooling and heating, star formation, and stellar feedback from radiation, stellar winds, and SNe. We refer the reader to M19 for more details of SMUGGLE. Here we highlight the sub-grid implementations that are relevant to this work. For star formation, star particles form from cold ($T<100\,\mathrm{K}$), dense ($n>100\,\mathrm{cm}^{-3}$), and self-gravitating molecular gas cells at a rate of:
\begin{equation}
\dot{M}_{\star,\mathrm{cell}} =
\epsilon_{\rm ff} \frac{M_\mathrm{gas,cell}}{t_{\mathrm{dyn,cell}}},
\end{equation} 
where $\dot{M}_{\star,\mathrm{cell}}$, $\epsilon_{\mathrm{ff}}$, and $t_{\mathrm{dyn,cell}}=\sqrt{\frac{3\pi}{32G \rho_\mathrm{cell}}}$ are the star formation rate, SFE per free-fall time, and gas free-fall time at the cell scale, respectively. Here, $\rho_\mathrm{cell}$ is the density of a gas cell. The initial conditions of the simulations were analog of a Milky Way-sized galaxy, which had a total mass of $\sim 1.6 \times 10^{12} M_{\odot}$ and contained a stellar bulge and disc, a gaseous disc, and a dark matter halo. The total mass of the gaseous disk was approximately $ \sim 9 \times 10^{9} M_{\odot}$, with a mass resolution of $1.4 \times 10^{3} M_{\odot}$ per gas cell. The stellar bulge had a total mass of $1.5 \times 10^{10} M_{\odot}$ , resolved at $2.3 \times 10^{3} M_{\odot}$ per stellar cell. Similarly, the stellar disk had a total mass of $4.73 \times 10^{10} M_{\odot}$ with a mass resolution of $1.9 \times 10^{3} M_{\odot}$ per stellar cell. We used adaptive gravitational softening for gas cells with a minimum softening length reaching $3.6$ pc, high enough to resolve the individual GMCs in the simulations. We adopted high-cadence snapshot write-out for every $1$ Myr to trace the temporal evolution of the clouds. The cloud tracking algorithm is elaborated further in Section~\ref{sec:establishtree} and Section~\ref{sec:calculateSFE}.

The whole suite contain six different model variations with different combinations of stellar feedback channels and star formation efficiency, namely SFE1, SFE10, SFE100, Rad, SN, and Nofeed. The SFE1 is the fiducial run with $\epsilon_{\mathrm{ff}}= 0.01$ and all stellar feedback channels. The SFE10 and SFE100 runs are the same as SFE1, but with $\epsilon_{\mathrm{ff}}= 0.1$ and $\epsilon_{\mathrm{ff}}= 1$, respectively. The Nofeed, Rad, and SN runs have the same $\epsilon_{\mathrm{ff}}$ as SFE1, but with different feedback methods: Nofeed run turns off all stellar feedback channels; Rad run includes only early feedback\footnote[1]{We note that ``early feedback'' includes both the stellar winds and radiative feedback, and we use the label ``Rad'' for the early feedback-only run because the radiative feedback always dominates the early feedback.}; SN run includes only SN feedback. We summarize the configuration of all model variations in Table~\ref{tab:modelparameters} and refer the reader to L20 for a detailed description of the numerical setup of the simulations and the parameters of the subgrid physics.

\begin{table}[h]
\centering
\caption{Summary of the six model variations analyzed in this work.}
\label{tab:modelparameters}

\begin{tabular}{lccc}
\hline
\hline
Name & $\epsilon_{\mathrm{ff}}$ &  SN feedback & Early feedback \\ 
\hline
SFE1       & $0.01$    & Yes           & Yes            \\
SFE10      & $0.1$    & Yes            & Yes            \\
SFE100     & $1$    & Yes            & Yes            \\
Rad        & $0.01$    & No           & Yes            \\
SN         & $0.01$    & Yes          & No            \\
Nofeed     & $0.01$    & No          & No            \\ 
\hline
\end{tabular}
\end{table}

\subsection{Identifying GMCs}\label{sec:iden_method}

To capture the life cycle of GMCs in the simulated galaxies, we first identified individual GMCs in all snapshots for all six simulations. We used {\sc CloudPhinder}\footnote[2]{\url{https://github.com/mikegrudic/CloudPhinder}}, which starts from local density peaks in galaxies and then identifies the largest self-gravitating gaseous structures by searching for neighboring cells. We set the volumetric density threshold to $50\,\mathrm{cm}^{-3}$,  similar to that of previous work \citep[e.g.,][]{Tasker_2009, Grisdale_2018MNRAS,2024MNRAS.534..215F:Fotopoulou}, and only considered gas cells denser than this threshold to find candidate clumps. We set a fairly tolerant criterion for the virial parameter $\alpha \leq \alpha_{\mathrm{max}}=10$ as we aimed to track clouds throughout their entire life cycles. We define the virial parameter of a given GMC as
\begin{equation}
\alpha = \frac{2 (E_{\mathrm{kin}} + E_{\mathrm{th}})}{|E_{\mathrm{grav}}|}.
\end{equation}
Here, $E_{\mathrm{kin}}$ and $E_{\mathrm{th}}$ are the kinetic and thermal energy of the cloud (measured in the cloud’s center-of-mass frame), and $E_{\mathrm{grav}}$ is the gravitational potential energy. The energies were computed as follows:
\begin{align}
&E_{\mathrm{kin}} = \sum_i \frac{1}{2} m_i |\vec{v_i} - \vec{v_c}|^2, \\
&E_{\mathrm{th}} = \sum_i m_i u_i, \\
&E_{\mathrm{grav}} = -\frac{1}{2} \sum_{i \neq j} \frac{G m_i m_j}{|\vec{r_i} - \vec{r_j}|},
\end{align}
where $m_i$, $u_i$, $\vec{v_i}$, and $\vec{r_i}$ are the mass, internal energy per unit mass, velocity, and position of the $i$-th gas cell, respectively, and $\vec{v_c}$ is the cloud’s center-of-mass velocity. $m_j$ and $\bm{r_j}$ are the mass and position of the $j$-th gas cell. To ensure that the identified GMCs are well resolved, we discarded clouds that contain fewer than ten gas cells\footnote[3]{We tested a threshold of up to $50$ cells and found that our main results—including cloud properties and lifetimes—remain robust. The only notable difference is the exclusion of some low-mass clouds at higher cell-number thresholds.}.

To understand the physical properties in our GMC population and better compare the identified GMCs with observations and other theoretical work, we define effective radius $R_{\mathrm{eff}}$ and velocity dispersion $\sigma_v$ of GMCs as 
\begin{equation*}
\begin{split}
    &R_{\mathrm{eff}}=\sqrt{\frac{5}{3}\frac{\sum (m_i r^2_i)}{\sum m_i}},\\
    &\sigma_v=\sqrt{ \frac{\sum [m_i (|\vec{v_i}-\vec{v_c}|^2+c^2_{s,i})]}   {\sum m_i} } ,
\end{split}
\end{equation*}
where $c_{s,i}$ is the sound speed of of the $i$-th gas cell.

\subsection{Constructing GMC evolution network}
\label{sec:establishtree}

After we identified GMCs across all epochs, we proceeded to track their temporal evolution and constructed an evolutionary network for each cloud. Similarly to \citet[][hereafter J21]{Jeffreson_2021MNRAS}, we represented each identified cloud as a ``node'' that could connect with other nodes via ``edges''. We stress that each node represented a GMC identified at a specific snapshot. It was a bundle of information summarizing a GMC’s state at a given time. Edges were conceptual links between nodes across adjacent snapshots. Edges existed purely in the network topology and did not correspond to physical boundaries or geometrical features of the clouds. Unlike J21, which tracked GMCs using only the $x$ and $y$ coordinates of gas cells (essentially a two-dimensional approach), we used the full three-dimensional spatial information $(x,y,z)$ of the gas elements to determine connections between clouds identified in consecutive snapshots. To establish links between clouds across snapshots, we utilized the unique cell IDs of gas elements. Although a small fraction of cells might change their IDs across different epochs due to mesh refinement or de-refinement in \textsc{arepo}, we found that the majority retained consistent IDs when the time interval between snapshots was short, i.e., $1$ Myr. This consistency enabled the matching of gas elements across snapshots and helped identify connections between GMCs. Specifically, two clouds, $A$ and $B$, were linked if they shared more than one common gas cell ID. This approach avoided assumptions about GMC movement trajectories required in a two-dimensional method and enabled more accurate tracking in three dimensions. To exclude marginal connections, we applied an additional criterion: the total mass of shared gas elements, $M_\mathrm{common}$, must exceed one percent of the mass of both clouds $A$ and $B$. Otherwise, the link between clouds $A$ and $B$ was discarded. After processing all clouds across snapshots, we produced a cloud evolution network consisting of nodes and edges for each simulation.

To further extract the lifespan of different clouds, we navigated through the network to construct cloud evolution trees, adhering to the following rules:

\begin{enumerate}

    \item Cloud nodes without any progenitor are root nodes, serving as the starting point for initiating new evolution trees. Every tree starting from its root node then goes along the network through time-forward edges among the cloud nodes.
    \item When a node finds no successor, the tree terminates at this point.
    \item If the node has only one successor in the network, the evolution tree proceeds directly to this node.
    \item For nodes with multiple successors, we calculate the mass fraction of common gas elements for each successor to determine the most probable path: For cloud $A$ at $t_1$ having $k$ successors at $t_2$, namely $\{B_1, B_2,..., B_k\}$, we record the mass fraction of the total common gas elements between cloud $A$ and each cloud $B_i (1\leq i\leq k)$ as $M_{A, B_i}$. Then we define the normalized probability for each path choice $A \xrightarrow{} B_i$ as
    \begin{equation*}
        P(A \rightarrow B_i)=\frac{M_{A,B_i}}{\sum_{1\leq i\leq k} M_{A,B_i}}
    \end{equation*}
The definition of $P(A \rightarrow B_i)$ remains consistent and applicable in the one-successor cases as $P(A \rightarrow B)=1$ when $k=1$.

\end{enumerate}

As a result, each root node yielded multiple potential trajectories originating from possible path choices within the multi-successor nodes. By multiplying the probabilities associated with each chosen path at each decision point, we calculated the probability for all trajectories emerging from each root node. 
Specifically, for the tree traversing the node path $X_0 \rightarrow X_1 \rightarrow X_2 \rightarrow ... \rightarrow X_n$, we derived the probability as 
\begin{equation}
    P(X_0 \rightarrow X_1 \rightarrow X_2 \rightarrow ... \rightarrow X_n)= \prod_{0}^{n-1} P(X_{i} \rightarrow X_{i+1}).
\end{equation} 
Then, we stored the path probabilities for all possible evolution trees in our cloud network, based on which the SFE will be calculated in Section~\ref{sec:calculateSFE} and the distribution of the cloud lifetimes will be derived in Section~\ref{sec:lifetimefit}. Note that our cloud evolution trees were different from those from the J21 approach in two key aspects: \begin{enumerate}
  \item The network in J21 considered any cloud with more descendants than progenitors as root nodes, which essentially treated all split clouds as new clouds. Here we started a new cloud evolution tree only when the cloud strictly had no progenitor. In this way, we could always capture the life cycles of GMCs from very early stages and follow their complete evolution.
  \item The network in J21 gave equivalent path probabilities for all descendants in every multiple-descendant cloud, regardless of how much gaseous mass flowed toward each descendant. Here we provided different weights to different descendants according to how much gaseous mass was flowing from the progenitor to each descendant. Thus, we focused on the majority of gaseous mass flow throughout cloud evolution trees.
\end{enumerate}

\subsection{Calculating GMC lifetimes and star formation efficiency}
\label{sec:calculateSFE}

The lifetime of GMCs, which contains information on how fast molecular gas is consumed in galaxies, is crucial for understanding galactic star formation efficiency. While individual cloud lifetimes could, in principle, be estimated by tracking each cloud from its formation to dissolution, due to the complex mass assembly history of clouds in different environments, we used a Monte Carlo (MC) sampling method with probability-weighted tree paths (as described in Section \ref{sec:establishtree}) to give a simple characterization of the cloud population across different simulation runs.

In each MC iteration, we set one random walker per root node and initialized the lifetime as $0$ Myr when the walker starts from the root node. Once a walker proceeded to the next node, i.e., it passed through one edge between two nodes, the lifetime of the current cloud tree increased by $1$ Myr, which corresponded to the temporal resolution of the constructed cloud network. In cases where a node had multiple descendants, the walker randomly chose one descendant to proceed based on path probabilities calculated in Section \ref{sec:establishtree}. To fully explore the tree hierarchy, we performed $1000$ MC iterations to reach convergence for all model variations.

To associate the star formation activities (star particles) with the cloud life cycles (cloud merger trees), we proposed the following approach. First, we defined a "main branch" for each root node, which referred to the path with the highest probability among all trajectories originating from that root node. Then star particles were matched to the main branches. For each node in any main branch, we recorded all gaseous cells' IDs from the cloud of this node and checked whether any of these IDs appeared in star particles in the next simulation snapshot. If a star particle in the next snapshot shared an ID with one of the gas cells in this node, we identified that star particle as having formed from this node. We could associate most star particles and cloud merger trees well via such an approach. However, a corner case arised when a gas cell rapidly moved from cloud $A$ to cloud $B$ within the snapshot interval $\sim 1\,\mathrm{Myr}$, became a star particle within $B$, and was incorrectly attributed to cloud $A$. To address this issue, we added a distance limit: The matched star particle must be located within $10$ times the effective radius ($10\times R_\mathrm{eff}$) of its host cloud. We confirmed that such cases were rare and that all results in this study remained unchanged if the distance limit was set to $5\times R_\mathrm{eff}$ instead of $10\times R_\mathrm{eff}$ as the distance limit.

After star particles and main branches were matched, we calculated the integrated star formation efficiency of each main branch as
\begin{equation}\label{eq:intSFE}
    \varepsilon_{\mathrm{int}}=\frac{ M_{\star}(t=t_{\mathrm{term}})}{ M_{\mathrm{max,baryon}} },
\end{equation}
where the maximum baryonic mass $M_{\mathrm{max,baryon}}=\mathrm{MAX\{M_{\mathrm{baryon}}(t)\}}$, and $M_{\mathrm{baryon}}(t)=M_{\mathrm{gas}}(t)+M_{\star}(t)$ is the baryonic mass (both gas and stars) of the cloud as a function of time. $t_{\mathrm{term}}$ means the moment that the cloud ends its life. For later discussion, here we also define $\Sigma_{\mathrm{gas}}(t)=\frac{M_{\mathrm{gas}}(t)}{4 \pi R^2_\mathrm{eff}(t)}$, and $\Sigma_{\mathrm{max}}=\frac{M_{\mathrm{max,baryon}}}{4 \pi R^2_\mathrm{eff}(t_\mathrm{mb})}$, where $t_\mathrm{mb}$ refers to the moment that the baryonic mass of a cloud reaches the maximum throughout its lifetime.

While $\varepsilon_{\mathrm{int}}$ reflects the overall fraction of gas converted to stars during the whole life cycle of each cloud, it is also theoretically interesting to estimate the instantaneous star formation efficiency per free-fall time: 
\begin{equation}\label{eq:SFEff}
    \varepsilon_{\mathrm{ff}(t)}=\frac{\dot M_{\star}(t) t_{\mathrm{ff}}(t)}{M_{\mathrm{gas}}(t)},
\end{equation}
where $t_{\mathrm{ff}}=\sqrt{3\pi/32G\bar{\rho}}$ is the gravitational free-fall timescale and the $\bar{\rho}$ is the average density over the volume of the cloud. Given that we determined the evolution tree at the finite time interval, the $\dot{M_\star}(t)$ was derived as
\begin{equation}
    \dot{M_\star}(t)=\frac{{M_\star}(t+\Delta t)-{M_\star}(t)}{\Delta t},
\end{equation}
where $\Delta t=1 \mathrm{Myr}$. We highlight that $\varepsilon_{\mathrm{ff}}$ represents the cloud-scale star formation efficiency per free-fall time, derived from cloud merger trees. This is distinct from the cell-scale star formation efficiency per free-fall time, $\epsilon_{\mathrm{ff}}$, which is a simulation runtime parameter described in Section \ref{sec:sim}.

\section{Results}\label{sec:results}

\begin{figure*}
\includegraphics[width=1\textwidth]{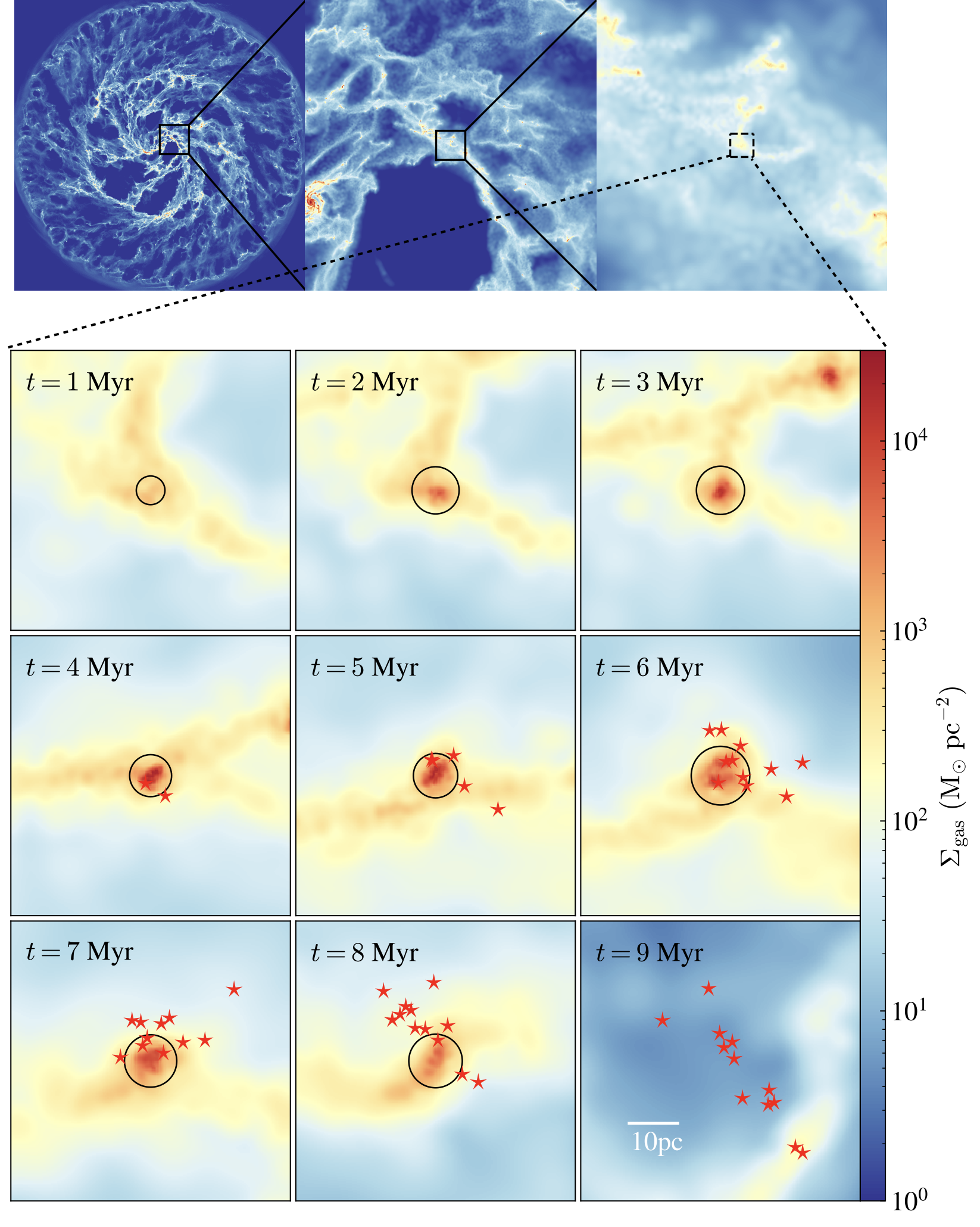}
\vspace{0mm}
\caption{Projections on the $x-y$ plane (face-on) for the gas surface density across the life cycle of one example GMC. The upper panel with three sub-panels shows respectively physical scales of $50$, $5$, and $0.5$ kpc on a side and illustrate the GMC's position within the simulated disk galaxy at its birth. The bottom nine panels follow the evolution of this GMC at different ages as shown on the upper left of each panel. All star particles that form in this GMC are denoted with red stars. The black circle centered at the center-of-mass of the cloud on each panel indicates is the effective radius of the cloud at that time.}
  \label{fig:demo_track}
\end{figure*}

\subsection{Visual impression of one GMC evolution}
\label{sec:visualization}

In total, we constructed $3.9 \times 10^4$, $9.0 \times 10^3$, $9.1 \times 10^1$, $8.3 \times 10^4$, $3.1 \times 10^4$, $1.4 \times 10^4$ cloud merger trees respectively in the SFE1, SFE10, SFE100, Rad, SN, and Nofeed runs. In this section, we show one representative tree in the fiducial run as a case study to demonstrate how our algorithm successfully tracks the evolution of one single GMC in the simulated galaxies. In the upper panel of Fig.~\ref{fig:demo_track}, the structure of one GMC identified in the fiducial run has been zoomed in from the large-scale galactic environment. The lower panel shows the evolutionary track of its life cycle. In the very early stages when we first identify the structure as the root node of the evolution tree ($1-2$~$\mathrm{Myr}$), the size ($\sim 3$~$\mathrm{pc}$), total gas mass ($\sim 10^{4}$~$M_{\odot}$) and gas surface density ($\sim 10^{3}$~$M_{\odot}\cdot \mathrm{pc}^{-2}$) of the cloud are small. During $2-4$~$\mathrm{Myr}$, as the cloud accretes gas from its surroundings, its mass and radius increase continuously. At $\sim 4$~$\mathrm{Myr}$, the cloud has gaseous mass $\sim 2 \times 10^{5}$~$M_{\odot}$, a value close to the maximum mass reached during its life cycle, and virial parameter $\alpha\sim 0.6$, therefore starts to form stars. At $\sim 6$~$\mathrm{Myr}$, the stellar mass within the GMC reaches its maximum $\sim 2 \times 10^{4}$~$M_{\odot}$. As young stars disturb the cloud through early feedback mechanisms, the virial parameter $\alpha$ increases slightly over unity. Finally, a considerable number of stars disperses the cloud at $\sim 8-9$~$\mathrm{Myr}$ until we lose track of the cloud. Quantitatively, the maximum baryonic mass over the life cycle of the cloud is $\sim 2.2 \times 10^{5}$~$M_{\odot}$ and the integrated SFE $\varepsilon_{\mathrm{int}}$ is approximately $9\%$ according to equation~(\ref*{eq:intSFE}).

\begin{figure}
\includegraphics[width=1\columnwidth]{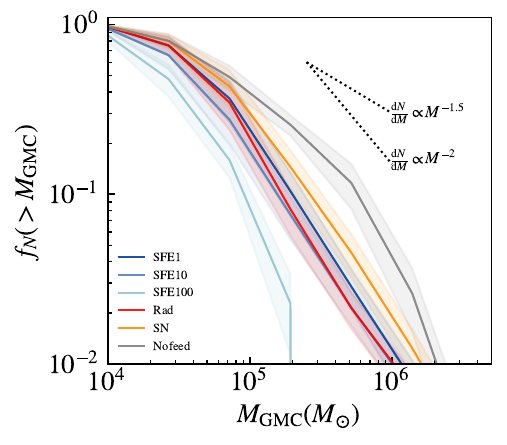}
\vspace{0mm}
\caption{Cumulative distribution of GMC mass ($M_{\mathrm{GMC}}$) in different model variations: SFE1 (royal blue), SFE10 (sky blue), SFE100 (light blue), Rad (red), SN (dark orange), and Nofeed (gray). Each solid line gives the median value of the cumulative mass function and the shaded areas enclose the $25 \% - 75 \%$ percentile across the time span $0.4-1$ Gyr. We also show the slopes of $1.5$ and $2$ as dashed black lines for reference. }
  \label{fig:cdf_mass}
\end{figure}

\begin{figure*}
\includegraphics[width=2\columnwidth]{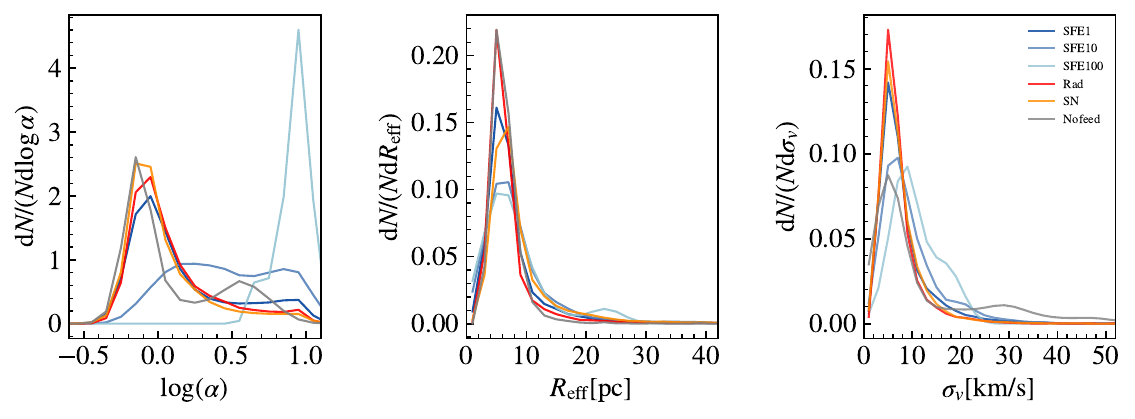}
\vspace{0mm}
\caption{From left to right: Normalized distribution of the virial parameter $\alpha$, effective radius $R_{\mathrm{eff}}$ and the velocity dispersion $\sigma_v$ of the GMCs in all model variations.}
  \label{fig:alphareffsigma}
\end{figure*}

\subsection{Physical properties of GMCs in different simulation runs}

For an overall view of the physical properties, especially the dynamical state, of our GMC population, we present the distribution of the mass ($M_{\mathrm{GMC}}$), virial parameter ($\alpha$), effective radius ($R_{\mathrm{eff}}$), and the velocity dispersion ($\sigma_v$) of GMCs in this section.

As described in Section~\ref{sec:iden_method}, we identified GMCs in each simulation snapshot from $0.4$ to $1$ Gyr. To avoid duplicated counting of the same clouds at different epochs of their life cycle, we collected the identified cloud sample from simulation snapshots at different epochs in time intervals of $50$ Myr for most runs except $\mathrm{SFE100}$ run. The choice of $50$ Myr is reasonable because it is longer than the lifetime of most clouds in our simulations. However, in the $\mathrm{SFE100}$, since lifetimes of the clouds run are extremely short (see Figure~\ref{fig:lifetime_multiruns}) and the number of identified GMCs is small, a time interval of $10$ Myr is used to enhance the sample size and reduce the scatter of the summary statistics.

In Fig. \ref{fig:cdf_mass}, the GMC mass distribution above $10^5 M_{\odot}$ in most runs except SFE100 and Nofeed has a power-law shape with a slope between -2 and -1.5, consistent with observations of Galactic GMCs \citep[e.g.,][]{Fukui_2010ARA&A}. The mass function of the GMCs in the SFE100 run shows a steeper slope, indicating the depletion of massive GMCs with a mass greater than $10^5 \Msun$. The extremely high $\epsilon_{\rm ff}$ in SFE100 leads to much faster consumption of cold gas and prevents clumps from growing into massive ones. It also leads to earlier and stronger stellar feedback that disperses the GMCs faster. Consequently, less massive GMCs in SFE100 tend to be weakly bound and have larger $\alpha$ (see Fig.~\ref{fig:alphareffsigma}). On the other extreme, in the Nofeed run, due to the lack of stellar feedback, there exists a larger number of massive GMCs with a shallower slope compared to the fiducial run.

In addition to mass, we also examine other key properties of GMCs that reflect their dynamical state, such as the virial parameter, effective radius, and velocity dispersion. The distribution of these quantities is shown in Figure~\ref{fig:alphareffsigma}. We find that the distribution of the virial parameter peaks at $\alpha<1$ for the SFE1, Rad, Nofeed, and SN runs. Given that we use a quite tolerant virial parameter threshold ($\alpha<10$) to identify clouds, it highlights that the dense molecular gas is gravitationally bound in nature in these runs. However, the situation changes in the SFE10 and SFE100 runs, where most of the clouds have virial parameters much larger than $1$. The unboundness of the detected structure is even more pronounced in the SFE100 run, where the distribution of GMCs exhibits a prominent peak around the threshold value, consistent with the fact that these runs lack massive clouds as shown above. For the Nofeed run, we notice that, besides the primary peak at $\alpha<1$, there exists a secondary peak around $\alpha \sim 3$. The bimodal distribution might indicate the existence of two different populations of GMCs, one of which is bound and the other unbound. Later in Section~\ref{res:sfe}, we will discuss more about this dichotomy and its subsequent evolution trend.

\begin{figure*}
\includegraphics[width=2\columnwidth]{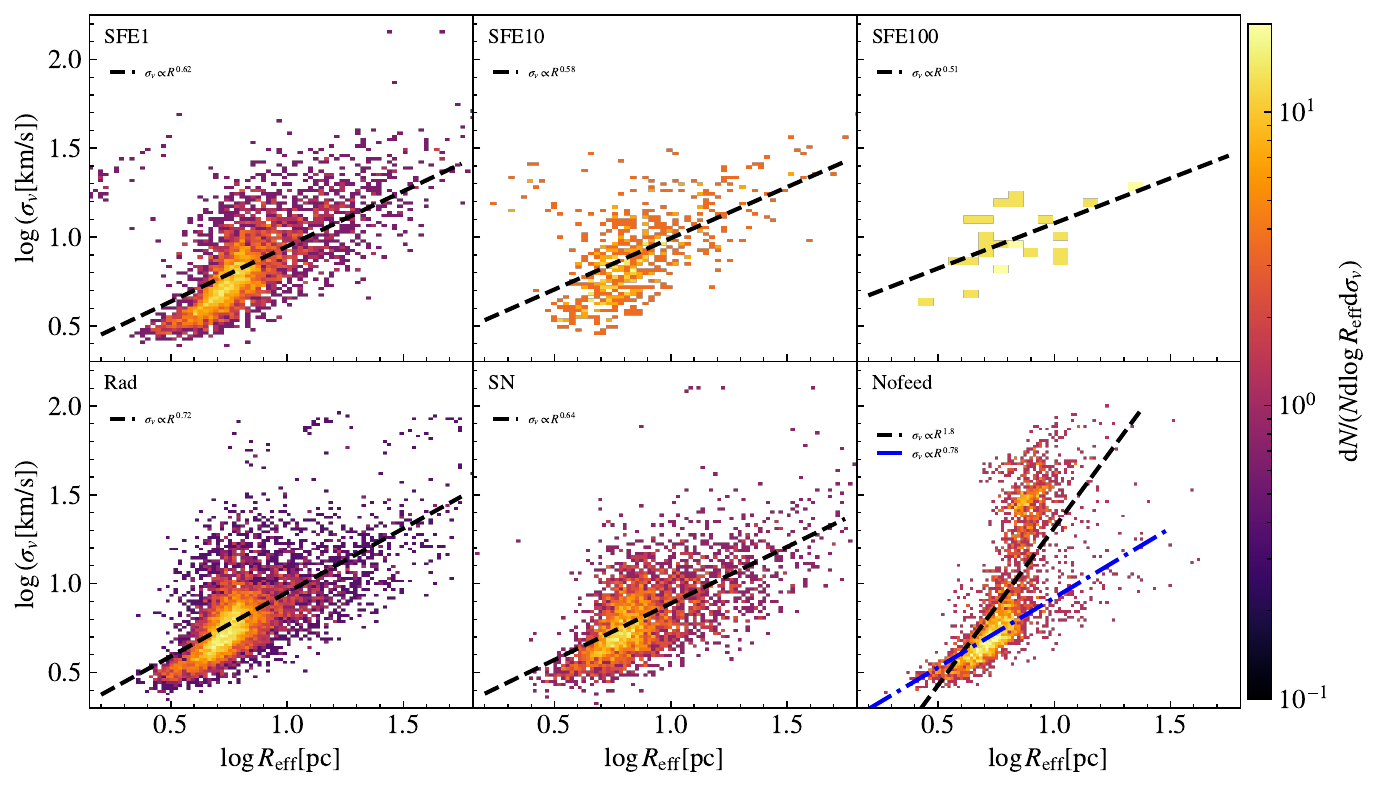}
\vspace{0mm}
\caption{2D histograms of the velocity dispersion ($\sigma_v$) and the effective radius ($R_{\mathrm{eff}}$) of the identified clouds in all model variations labeled at the upper left corner in each panel. We fit a Larson-like relation for each panel and the best-fit relation is over-plotted as dashed black lines. In the Nofeed run, the dashed blue line shows an additional fit to the data limited to the $\sigma_v<1 \,\mathrm{km}\,\mathrm{s}^{-1}$ region. }
  \label{fig:sigmar}
\end{figure*}

In observation of local GMCs, there exists a well-known empirical relation, the so-called Larson's Law, between the size and velocity dispersion of the clouds over a wide range of cloud masses and sizes \citep{Larson1981MNRAS}: $\sigma_v \propto R^b$. To investigate whether the GMC populations in this work follow similar a trend, we show the correlation between the velocity dispersion of GMCs $\sigma_v$ and $R_{\mathrm{eff}}$ in Fig. \ref{fig:sigmar}. The index $b \sim 0.62$ in the fiducial run is very close to the best-fitting for MW GMCs as $b=0.63 \pm 0.30$ \citep{MivilleDesch_2017ApJ}, suggesting that the cold ISM properties in the simulation are realistic and the cloud identification procedure is reliable. Other runs with different $\epsff$ and feedback mechanisms show similar slopes from 0.51 to 0.72, all within the uncertainties level of the observations. In the Nofeed run, however, we found a much steeper slope with $b\sim1.8$ caused by a large number of GMCs with $R_\mathrm{eff}$ all close to $10 \mathrm{pc}$ but having a large variation on $\sigma_v$. Since there is neither photoionization, radiative pressure nor SN feedback in the Nofeed run, those GMCs in the high velocity dispersion regime ($\sigma_v>10 \mathrm{km}/\mathrm{s}$) can only be significantly disturbed by the galactic dynamics. The existence of these high-$\sigma_v$ GMCs is also consistent with the minor peak in the distribution of virial parameter for the Nofeed run in Figure~\ref{fig:alphareffsigma}. While we will further discuss the effects of galactic dynamics on GMCs in Section~\ref{res:sfe}, here we further exclude those GMCs with $\sigma_v > 10 \mathrm{km}/\mathrm{s}$ and get the slope $b \sim 0.78$ for those GMCs less disturbed by galactic dynamics.

\subsection{GMC lifetime distribution}\label{sec:lifetimefit}

\begin{figure}
\includegraphics[width=\columnwidth]{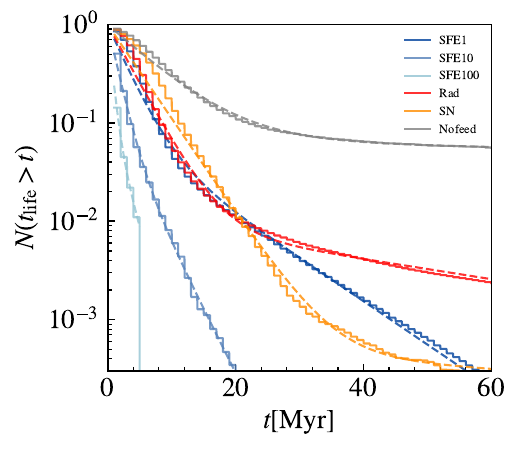}
\vspace{0mm}
\caption{Cumulative distribution of the cloud lifetimes (solid) and the best-fit to a double exponential distribution (dashed) for all model variations.}
  \label{fig:lifetime_multiruns}
\end{figure}

In Figure~\ref{fig:lifetime_multiruns}, we show the cumulative distribution of GMC lifetimes in all model variations. Compared with the fiducial run, the cumulative lifetime distribution of GMCs in the SFE10 run exhibits a steeper slope, and this trend becomes even more pronounced in the SFE100 run. On the one hand, a higher star formation efficiency per free-fall time leads to the faster consumption of gas, so GMCs live shorter in simulations with higher $\epsilon_{\rm ff}$. On the other hand, more efficient star formation makes the conversion from gas to stars earlier and therefore brings stellar feedback earlier before more gas is accreted, which also results in a more rapid termination of GMC life cycles in SFE10 and SFE100 runs.

We then focus on the effects on cloud lifetimes from different channels of stellar feedback by comparing cloud lifetimes in the Rad run and the SN run with the fiducial run. In the Rad run, the cloud lifetime distribution follows almost the same slope as in the fiducial run for short-lived clouds with $t_\mathrm{life}<20\,\mathrm{Myr}$, suggesting that their evolution is mostly controlled by early feedback rather than SN feedback. However, for $t_\mathrm{life}>20\,\mathrm{Myr}$, Rad shows an extended tail with a shallower slope compared with the fiducial run. This difference directly shows that many long-lived clouds cannot be disrupted by early feedback alone but have to wait for SN to quench the star formation activities. As we discuss in Section~\ref{sec:mt_relation} (see also Fig. ~\ref{fig:mle_mt_fit}), these long-lived clouds are systematically more massive and have higher gas surface density, than the short-lived ones. Therefore, the relatively mild photoionization and radiative pressure are too weak to terminate the lives of such massive clouds in the lack of sudden momentum and energy feedback from the supernova. On the opposite, we report the shallower slope at the low lifetime end in the SN run. For these short-lived clouds, a supernova needs more time than the cloud's lifetime to take effect so there is nearly no feedback during the life cycles of those short-lived clouds.

For GMC lifetimes in different simulations, J21 gives a characteristic index $\tau$ to describe the GMC lifetime distribution within a certain sample. They assume an exponential distribution and extract the characteristic cloud lifetime from the power-law index. We, however, find a turnover in the cumulative distribution of cloud lifetimes and we consider the distribution of the cloud lifetimes as the combination of two independent components, one of which shows intrinsically longer characteristic lifetimes than the other one. Here we define $\tau_1,\tau_2$ representing the characteristic lifetimes for the two populations of the clouds, and their cloud number fractions are noted as $\eta$ and  $ (1-\eta)$ respectively. Therefore, the distribution $N(t_{\mathrm{life}}>t)$ can be described by
\begin{equation}\label{eq:1}
    N(t) = N_0 [\eta \cdot e^{-\frac{t}{\tau_1}} + (1-\eta) \cdot e^{-\frac{t}{\tau_2}}],
\end{equation}
where $N_0$ is the normalization factor and is equivalent to the total number of clouds in each run.
We then fit the distribution in the log-linear space and present the best values of the parameters in Table~\ref{tab:fitresults}. In Fig.~\ref{fig:lifetime_multiruns} we find that the fitting curves agree with the data well among all model variations.

\begin{table}[h]
\centering
\caption{Best-fitting parameters for the distribution of GMC lifetimes in all simulations.}
\label{tab:fitresults}

\begin{tabular}{lccc}
\hline
\hline
Name       & $\eta$                  & $\tau_1$ (Myr) & $\tau_2$ (Myr) \\ 
\hline
SFE1       & $8.9 \times 10^{-2}$    & 9.8            & 2.8            \\
SFE10      & $1.1 \times 10^{-1}$    & 3.3            & 1.4            \\
SFE100     & $1.2 \times 10^{-1}$    & 1.9            & 0.6            \\
Rad        & $1.1 \times 10^{-2}$    & 42.4           & 3.6            \\
SN         & $5.9 \times 10^{-4}$    & 101.2          & 4.5            \\
Nofeed     & $6.9 \times 10^{-2}$    & 306.8          & 7.1            \\ 
\hline
\end{tabular}
\end{table}

\begin{figure}
\includegraphics[width=1\columnwidth]{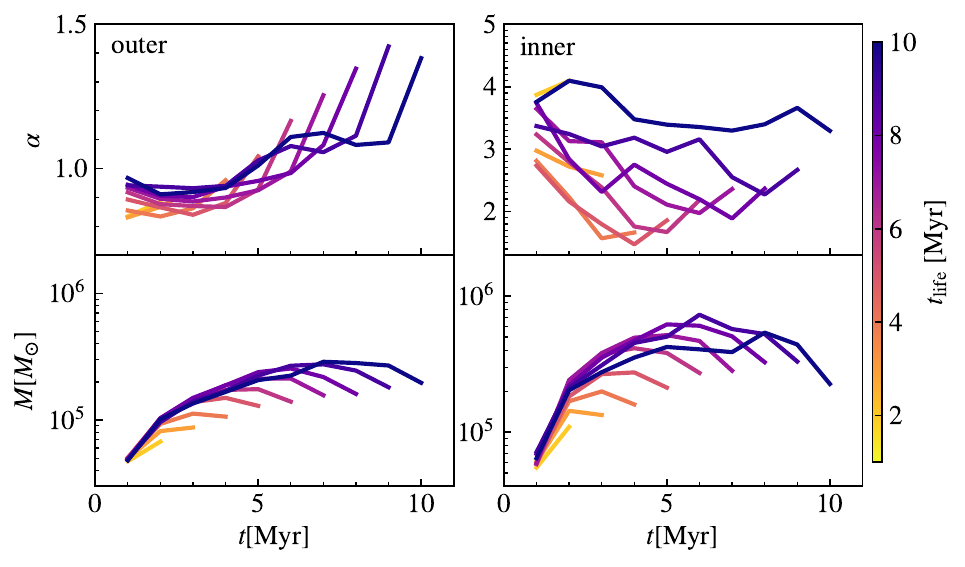}
\vspace{0mm}
\caption{Temporal evolution of the virial parameter $\alpha$ and the gaseous mass $M$ of the GMCs in the outer (left, $>2\,\rm kpc$) and inner (right, $<2\,\rm kpc$) galactic disk in the SFE1 run. Each line shows the evolution of the median value among clouds within the same lifetime bin indicated by different colors as in the colorbar.}
  \label{fig:alphaMR}
\end{figure}

To further address the physical nature of the two populations of clouds, we then separate the clouds in the fiducial run into two groups based on the distance of the clouds to the galactic center and plot the temporal evolution of $\alpha$ and $M$ in each group respectively. In Fig.~\ref{fig:alphaMR}, we find a distinct difference in the evolution of these two groups. The clouds beyond $2$ kpc from the galactic center go through life cycles as follows: The gaseous mass of these clouds is relatively low compared to the mass of clouds in the inner galactic region, e.g., $\sim 10^{4}$--$10^{5}\,M_{\odot}$, and they are still accreting gas from the galactic environment. The virial parameters of these clouds generally start from a low value indicating the gravitationally bound state and then decrease during the accretion process. The gaseous masses correspondingly grow in this accreting stage. When it comes to the $1-2\, \mathrm{Myr}$ before the termination of clouds' life cycles, the clouds are massive enough to collapse. After collapsing to form stars, clouds tend to be more gravitationally unbound due to the feedback mechanism and the virial parameters reach a much larger value. For the group of clouds closer to the galactic center, we however, see the quite different and significantly more disturbed trend of the virial parameters as the function of time. Compared with clouds in the inner region, these clouds are generally more massive and originate from unbound clumps. During their life cycles, they may remain marginally unbound due to the strong galactic scale effects on cloud dynamics by a combination of galactic shear and external pressure. The spatial separation is chosen as $2$ kpc here but we have checked that the above two modes exist similarly if the spatial separation is chosen as $1$, $1.5$, or $2.5$ kpc. We have also verified that the gas dynamics of clouds in the Nofeed, Rad, and SN runs all exhibit a similar spatial distribution trend.

\subsection{Relation between GMC lifetime and maximum gaseous mass}\label{sec:mt_relation}

\begin{figure}
\includegraphics[width=\columnwidth]{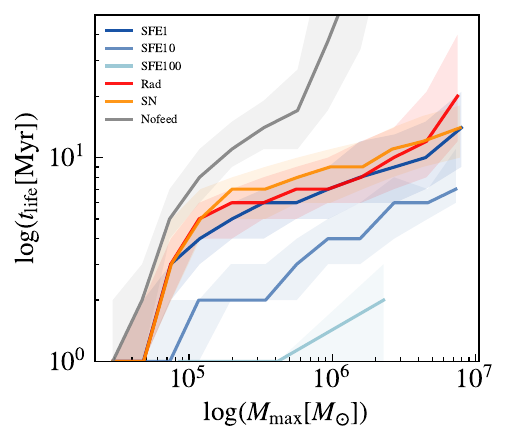}
\vspace{0mm}
\caption{Relation between the cloud lifetimes $t_{\mathrm{life}}$ and the maximum gaseous masses $M_\mathrm{max}$ for the clouds in all model variations. The solid lines show the median while shaded regions enclose the $25\%-75\%$ percentile of the lifetimes in each mass bin. }
\label{fig:mle_mt_fit}
\end{figure}

In the bottom left panel of Fig.~\ref{fig:alphaMR}, we observe that for clouds located outside the galaxy's central region, the accretion process continuously supplements their gas content throughout most of their life cycle. Then they get disrupted quickly in a couple of million years and eventually, we lose track of them. Inspired by such a trend, we then analyze how the maximum gas mass of a cloud relates to its lifetime in Fig. ~\ref{fig:mle_mt_fit}. The discrete distribution in the lifetimes is due to the finite temporal resolution of our cloud evolution tree of $1$ Myr. In the SFE100 run, owing to the temporal resolution of the evolutionary tree, we could barely resolve the life cycles of the low-mass clouds due to the rapid gas consumption via the star formation process in these clouds. Despite the model variations in different simulation runs, all the lines show a significant positive correlation between the lifetimes and the gaseous masses in general, in agreement with the continuous accreting state of the GMCs in their early life. The lines for the Rad and SN runs both rise faster and are slightly higher than the line for the SFE1 run, while the Nofeed run corresponds to the line with the steepest slope across $\sim 4\times 10^4 M_{\odot}$ to $\sim 5\times 10^6 M_{\odot}$. Such a trend is consistent with the fact that both early feedback and SN feedback contribute to dispersing clouds. In most runs except SFE100, we find a rapid increase in the lifetime of GMCs in the low mass regime ($< 10^5 M_{\odot}$). But then the slope between the lifetime and gaseous mass becomes shallower in the $M_\mathrm{max}> 10^5 M_{\odot}$ regime. On the one hand, GMCs in their late lives may accrete gas from surrounding ISM not as efficiently as in their early lives. On the other hand, the star formation process continuously consumes the gas of GMCs. The combined effect of the above causes the different slopes between the lifetime and gaseous mass in different regimes.

\begin{figure*}
\includegraphics[width=2\columnwidth]{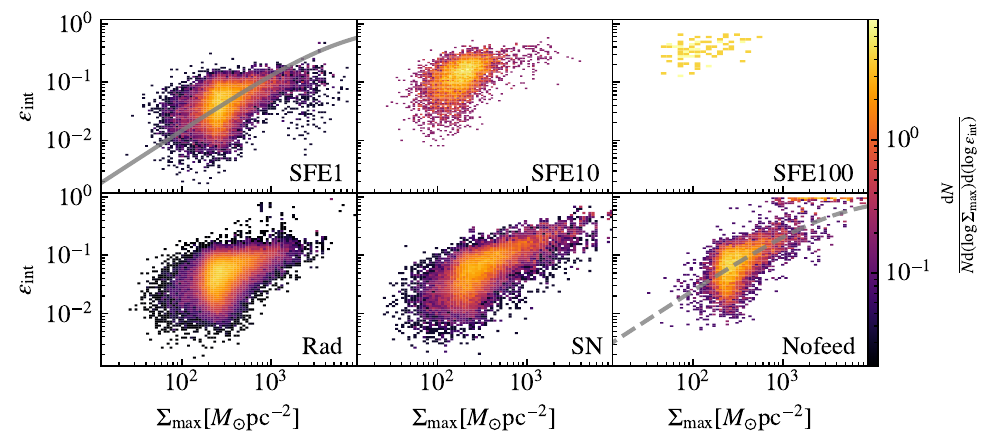}
\vspace{0mm}
\caption{2D histograms of the integrated star formation efficiency $\varepsilon_{\mathrm{int}}$ and the maximum baryonic surface density $\Sigma_\mathrm{max}$ of the clouds in all model variations labeled in the bottom right corner of each panel. Clouds with an integrated star formation efficiency equal to zero are excluded. The dashed line is the best-fitting model for Nofeed run based on equation (\ref{eq:SFEbyGS}) considering only the effects of galactic shear while the solid line is the best-fitting model to equation (\ref{eq:SFEwithShear}) that considers both the galactic shear and the stellar feedback \citep{Grudic_2018MNRAS,Li_2019MNRAS}.} 
  \label{fig:epsilon_sigma}
\end{figure*}

\subsection{Star formation efficiency on cloud scales}\label{res:sfe}

Recently, it has been realized from many theoretical and numerical works \citep[e.g.,][]{Murray_2010ApJ,Grudic_2018MNRAS,Li_2019MNRAS} that there exists a positive correlation between integrated star formation efficiency and gas surface density of clouds:

\begin{equation}\label{eq:SFEbyGS}
    \varepsilon_{\mathrm{int}}=\left(1+\frac{\Sigma_{\mathrm{fb}}}{\Sigma_{\mathrm{max}}} \right)^{-1},
\end{equation}
where critical feedback density $\Sigma_{\mathrm{fb}}=\frac{\dot{p_{\star}}}{4 \pi G}$ and $\dot{ p_{\star}}$ is the time- and IMF-averaged rate of the specific momentum feedback strength from massive stars. The physical origin of this correlation is the force balance between the self-gravity of the cloud and momentum stellar feedback from massive stars \citep[e.g.,][]{Fall_2010ApJ, Li_2019MNRAS}.
This physical interpretation assumes that the cloud is in isolation without accreting material and without external forces from the ambient environment. Here we test this scenario using our galaxy simulations with resolved GMC populations.

Based on the method described in Section \ref{sec:calculateSFE}, we calculate the integrated SFE for each GMC along the main branch of the network. As we can see in Fig.~\ref{fig:epsilon_sigma}, in general, GMCs in simulated galaxies follow a similar positive correlation between $\Sigma_{\rm max}$ and $\varepsilon_{\rm int}$ for all model variations. In the fiducial run, the correlation between $\Sigma_{\rm max}$ and $\varepsilon_{\rm int}$ is qualitatively in agreement with the isolated cloud studies \citep[e.g.,][]{Grudic_2018MNRAS,Li_2019MNRAS} shown by equation~(\ref{eq:SFEbyGS}). Comparing the fiducial and the SFE10 and SFE100 runs, we find that increasing $\epsilon_\mathrm{ff}$ leads to faster depletion of high gas surface density clouds because of high star formation on the cell level and thus earlier and stronger feedback. For lower $\Sigma_{\mathrm{max}}$ clouds, higher $\epsilon_\mathrm{ff}$ leads to higher $\varepsilon_{\rm int}$ meaning that the cloud-scale integrated efficiency follows the same trend of the cell-level efficiency. This is consistent with the trend found in the cosmological simulations \citep{Agertz_2013ApJ, 2018ApJ...861..107L:Li, Oh_2020MNRAS, Nuezcastieyra_2021MNRAS}. It is worth emphasizing here that the SFE100 is the most extreme case in which only a sparse number of clouds are identified, making them hard to track.

While the general distribution of $\varepsilon_{\mathrm{int}}$ is roughly consistent with equation~(\ref{eq:SFEbyGS}) in the fiducial run, we would like to investigate how this relation varies when different channels of stellar feedback are used by comparing the GMC sample in the Rad, SN, and Nofeed runs. Unlike the fiducial run where stellar wind and radiation pressure from newly born stars prevent GMCs from collapsing to an extremely high density in the early stage of star formation, in the SN run a considerable number of GMCs manage to grow into extremely dense clumps with $\Sigma_{\mathrm{max}}$ even close to $10^4\,M_{\odot}\mathrm{pc}^{-2}$ due to the lack of early stellar feedback. Clouds that reach high surface densities and collapse to form new stars can remain bound and continue star formation in the absence of early feedback until eventually being dispersed by supernovae from the older stars.

In the Rad run, we report a similar distribution of the GMC candidates in the $\Sigma_{\mathrm{max}}$-$\varepsilon_{\mathrm{int}}$ plane to that of the fiducial run. However, for clouds with $\Sigma_{\mathrm{max}}>10^3 \;M_{\odot}\mathrm{pc}^{-2}$, there exists a small fraction of dense clouds with a significantly higher $\varepsilon_{\mathrm{int}}$ than those in the fiducial run. Since there is no SN feedback in the Rad run, once the cloud is dense enough, it is hard to be dispersed by early feedback. Without the large amount of energy and momentum ejected by any supernova explosion, such dense GMCs survive longer and keep on collapsing to form new stars, making the integrated star formation efficiency much higher than that in the fiducial run.

In the Nofeed run, as expected, a considerable number of clouds with $\Sigma_{\mathrm{max}}>10^3 \;M_{\odot} \mathrm{pc}^{-2}$ show extremely efficient star formation with the majority of the gaseous mass turning into stellar masses, i.e., reaching almost $100\%$ integrated star formation efficiency. Intriguingly, without any stellar feedback input, there are still a large number of clouds showing relatively low integrated star formation efficiency during their lifetime, especially for clouds with $\Sigma_{\mathrm{max}}<10^3 \;M_{\odot}\mathrm{pc}^{-2}$, and the positive correlation between $\Sigma_{\mathrm{max}}$ and $\varepsilon_{\mathrm{int}}$ still roughly holds. This suggests that other mechanisms, possibly galactic dynamics, act as an effective ``feedback source'' that is responsible for the disruption of the GMCs. 

Following this logic, we improve the model in equation~(\ref{eq:SFEbyGS}) by adding a galactic shear term that takes into account the environmental effects on the galactic scale. For simplicity, we introduce the critical shear surface density $\Sigma_{\mathrm{shear}}$ which effectively reflects how the galactic shear/tides suppress gravitational collapse. The equation~(\ref{eq:SFEbyGS}) then becomes
\begin{equation}\label{eq:SFEwithShear}
    \varepsilon_{\mathrm{int}}=\left(1+\frac{\Sigma_{\mathrm{fb}}+\Sigma_{\mathrm{shear}}}{\Sigma_{\mathrm{max}}}\right)^{-1}.
\end{equation}
In the Nofeed run, as $\Sigma_{\mathrm{fb}}=0$ we can independently fit $\Sigma_{\mathrm{shear}}$ within our cloud sample. The best-fit value of the critical shear density is $\Sigma_{\mathrm{shear}}=3.9 \times 10^3  M_{\odot}  \mathrm{pc}^{-2}$, which can be interpreted as a galaxy-wide average strength of the galactic shear and tides. Using this constant and fitting equation~(\ref{eq:SFEwithShear}) to the SFE1 samples, we obtain the critical feedback density as $\Sigma_{\mathrm{fb}}=2.8 \times 10^3 M_{\odot} \mathrm{pc}^{-2}$. Although $\Sigma_{\mathrm{shear}}$ is not a first-principle measurement of galactic environmental effects on GMCs, the main takeaway from the above fit is that external galactic environments play a similar, if not bigger, role than internal stellar feedback in shaping GMC life cycles.

Despite the general trend between $\varepsilon_{\mathrm{int}}$ and $\Sigma_{\mathrm{max}}$ can be roughly described by equation (\ref{eq:SFEwithShear}), we observe large scatters around the mean relation in Fig.~\ref{fig:epsilon_sigma}. This scatter is present in all model variations, including the Nofeed run. This scatter suggests that, besides $\Sigma_{\mathrm{max}}$, there exists other physical variables that affect the cloud-scale star formation efficiency. We find that the scatter depends on the cloud lifetime: for a given $\Sigma_{\mathrm{max}}$, clouds with longer lifetime tend to have larger $\varepsilon_{\mathrm{int}}$. We also find a weak relationship between the scattering and the strength of galactic tides across different locations and epochs. We will explore the environmental effects in detail in a follow-up paper (Deng et al. in prep.)

\subsection{Star formation efficiency per free-fall time in GMC life cycles}\label{res:sfeff}

\begin{figure}
\includegraphics[width=1\columnwidth]{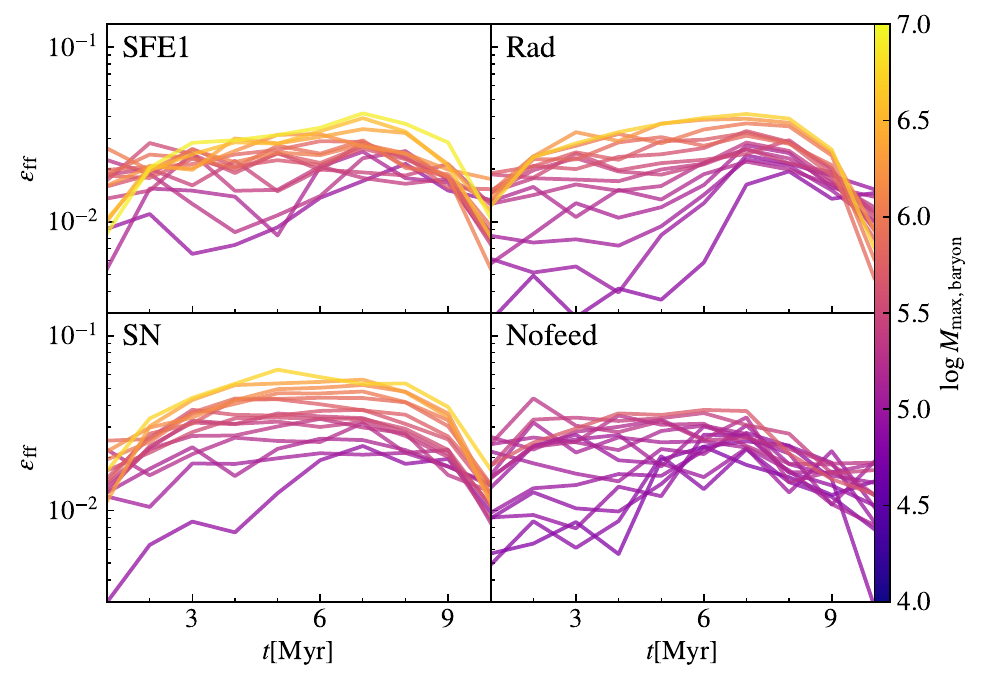}
\vspace{0mm}
\caption{Temporal evolution of the star formation efficiency per free-fall time $\varepsilon_{\mathrm{ff}}$ in the clouds that survive for $\sim10$ Myr in SFE1, Rad, SN, Nofeed runs. The cloud sample is separated into bins of different maximum baryonic mass $M_\mathrm{max,baryon}$, as indicated by different colors of the lines. In each bin of maximum baryonic mass, $\varepsilon_{\mathrm{ff}}$ and $M_\mathrm{max,baryon}$ are averaged respectively.}
  \label{fig:epsilon_t_evlove}
\end{figure}

To further follow the SFE in different stages within the life cycles of GMCs, we calculate the star formation efficiency per free-fall time $\varepsilon_{\mathrm{ff}}$ by equation~(\ref{eq:intSFE}) for all tracked GMCs in the SFE1, Rad, SN, and Nofeed runs. We then select all $t_\mathrm{life}=10\,\mathrm{Myr}$ GMCs among them and for each model variation, we divide $t_\mathrm{life}=10\,\mathrm{Myr}$ GMCs into $15$ equal-cloud number bins according to the maximum baryonic mass $M_{\mathrm{max,baryon}}$. Within each maximum baryonic mass bin, we average $\varepsilon_{\mathrm{ff}}$ among GMCs at every moment and get $\varepsilon_{\mathrm{ff}}$ as a function of time $t$ in Fig.~\ref{fig:epsilon_t_evlove}. Here we note that it is almost impossible to choose a fair value of GMC lifetime in this comparison of $4$ simulations with various feedback setups. On the one hand, we want to have sufficient clouds to be statistically meaningful, so we cannot choose a too large value for a lifetime. On the other hand, we do not want to choose a too short value to miss any physical effect in the late life of clouds such as supernovae and long-term dynamics. Here we choose $10$ Myr as a balance between these limits. In general, GMCs among all of Rad, SN, and Nofeed runs show an increasing $\varepsilon_{\mathrm{ff}}$ in the early stages and decreasing $\varepsilon_{\mathrm{ff}}$ in the last several million years of their life cycles. Such a trend is consistent with the results in \citet{Grudic_2019MNRAS}. Among these four model variations, GMCs with higher $M_{\mathrm{max,baryon}}$ exhibit higher $\varepsilon_{\mathrm{ff}}$. The high $\varepsilon_{\mathrm{ff}}$ (larger than $2\%$) in more massive GMCs can sustain for the entire life cycles of these GMCs, while less massive GMCs can have extremely low $\varepsilon_{\mathrm{ff}}$ (less than $2\%$) in the first few million years. Compared with both the SFE1 run and the Rad run, the star formation in the SN run is slightly more active across all mass ranges. Given that the selected timescale is $10$ Myr here, it is relatively short for the supernova to take effects thus the star formation in the SN run is mainly regulated by the galactic environment.

\section{Discussion}\label{sec:discussion}

\subsection{Comparisons to previous work}

During the past several years, thanks to the increasing power of supercomputers and a better understanding of the ISM physics in galaxy formation simulations, a few studies have investigated the GMC properties from large-scale galaxy simulations. Here we would like to compare our results with them. Regarding the physical properties of identified GMCs, such as mass, virial parameter, effective radius, velocity dispersion, and Larson's relations, GMCs identified in the Feedback ($3$D) case from \citet{Grisdale_2018MNRAS} in general exhibit similar distributions as shown in our fiducial run. The mass distribution of GMCs in the m12m case of \citet{Benincasa_2020MNRAS} is also similar to that in our fiducial run. \citet{Guszejnov_2020MNRAS} reports similar mass distribution of GMCs and similar Larson-like relation between $\sigma_v$ and $R_\mathrm{eff}$ in cosmological simulations.

In terms of GMC lifetimes, \citet{Richings_2016MNRAS}, \citet{Benincasa_2020MNRAS}, J21, and our work all find that GMC lifetimes range from a couple of million years to several tens of million years. While the authors do not explicitly discuss trends between GMC lifetime and mass, in \citet{Richings_2016MNRAS}, we note that their fiducial tracking scheme present a positive correlation between GMC lifetimes and masses, consistent with our conclusions. However, \citet{Benincasa_2020MNRAS} report limited dependence of the lifetime on the GMC mass and J21 proposes a negative correlation between GMC lifetimes and masses, which are contradictory to our findings on the positive correlation between GMC lifetimes and masses. In \citet{Benincasa_2020MNRAS}, they used a relatively simple tracking algorithm and only followed strictly gravitationally bound GMCs. This approach might overlook the early stages of the GMC life cycles and miss key accretion phases, which are crucial to establishing the positive correlation between GMC lifetimes and masses. As for J21, though they introduced a similar tree network approach to capture both the temporal evolution of clouds and their interactions like our work, here we highlight one significant difference between J21's and our GMC-tracking method, which may cause different results on the relation of GMC lifetimes and masses. Using the quasi-Lagrangian nature of {\sc arepo}, we recorded additional information about the gas mass flow between nodes, which provided physical guidelines for selecting paths among multiple branches. J21, on the contrary, assumed the same probabilities for branches, regardless of the gas mass flowing through them. In this way, the GMC tracking in J21 might be misled by statistically overweighting minor branches of GMC trees. A massive GMC may go through many splitting or merging processes within its lifetime, and thus produce many minor branches. According to their tracking algorithm, these minor branches shorten the characteristic lifetimes of massive GMCs. It should be mentioned that in the GRIFFIN project, only clouds with the longest lifetime can form the most massive star clusters \citep{Lahen_2020ApJ}, consistent with our findings here.

Regarding the star formation process, while many studies focus on isolated GMC simulations to explore local SFE, our approach captures the influence of galactic environments self-consistently, while still enabling investigations of local SFE. In Section \ref{res:sfe}, we notice that the relation between the cloud surface density and the SFE is in a similar form as found in the isolated GMC simulations \citep[e.g.,][]{Grudic_2018MNRAS, Li_2019MNRAS}. Moreover, the large variations in the local SFE are observed in all simulation runs, consistent with \citet{Grisdale_2019MNRAS}. Besides the fiducial run with the stellar feedback, we utilize the GMCs in the Nofeed run to quantitatively approximate the effect of the galactic shear on the local star formation efficiency with $\Sigma_{\mathrm{shear}}$. The best-fitting value of $\Sigma_{\mathrm{shear}}$ is comparable to the best-fitting value of $\Sigma_{\mathrm{fb}}$, suggesting the non-negligible role that the galactic shear plays in the star formation activity. We would like to point out that in \citet{Grudic_2018MNRAS} and \citet{Grudic_2023MNRAS}, while their best-fitting $\Sigma_{\mathrm{fb}}$ is close to our best-fitting value, an extra parameter namely $\epsilon_{\mathrm{max}}$ is artificially introduced to account for any possible additional suppression of star formation other than stellar feedback. This approximation may partly mimic the effect of the $\Sigma_{\mathrm{shear}}$ in our study. Similar to our work, \citet{Khullar_2024ApJ} also track life cycles of GMCs and study the GMC-scale SFE in galactic scale simulations, they get similar results on the importance of stellar feedback regulating star formation and also the large scatter in local SFE.

\subsection{Limitations and future perspectives}

We acknowledge several limitations of this work and outline potential directions for future research. First, while \textsc{arepo}, as a moving-mesh code, has a quasi-Lagrangian nature, it is not purely Lagrangian. The timescale for cell refinement and de-refinement is much shorter than the snapshot output cadence (i.e., $1\,\mathrm{Myr}$). Consequently, using cell IDs to track mass flow between GMCs across different epochs can lead to inaccuracies due to cell reconstruction during these intervals. To address this, future simulations will incorporate the Monte Carlo tracer particles \citep{Genel_2013MNRAS} to allow the mass flow of clouds to be tracked more accurately. Second, though we find that galactic shear has a significant effect on GMC-scale star formation, it is yet unclear how such an effect would vary in different galactic environments. To explore how various galactic environments might have different effects on GMC-scale star formation, a companion study (Deng et al., in prep.) is currently underway to explore the evolution of GMCs in dwarf galaxies, utilizing the solar-mass-resolution ISM model, RIGEL \citep{2024A&A...691A.231D:Deng}. 

\section{Summary}\label{sec:summary}

In this study, we tracked the temporal evolution of a large sample of GMCs within simulated Milky Way-sized galaxies using the hydrodynamic code \textsc{arepo} under the SMUGGLE galaxy formation framework. The high mass resolution ($\sim10^3 M_\odot$) and high-cadence output frequency ($1$ Myr) enable the identification of cloud candidates and the construction of evolutionary trees. These evolutionary trees successfully capture the full GMC life cycle, including formation, gas accretion and collapse, star and cluster formation, and cloud dispersal. To investigate how different physical processes influence GMC evolution and star formation on cloud scales, we explore a set of model variations with different star formation and stellar feedback parameters. Below, we summarize the key findings:

\begin{itemize}

    \item The identified GMCs show a strong correlation between their size and velocity dispersion $\sigma_v \propto R^b$, with $b$ ranging from $0.51$ to $0.73$ in most runs with stellar feedback, generally consistent with the observed MW GMCs. However, GMCs in the no feedback run (Nofeed) show a much larger value of $b\sim1.8$, significantly deviating from observations. This suggests that stellar feedback is crucial for regulating the statistical properties of the ISM turbulence in the galactic disk, thereby influencing the internal structure of clouds.

    \item GMC lifetimes vary significantly across model variations, ranging from a few to several tens of million years, suggesting a strong influence from the star formation and stellar feedback subgrid models. Higher $\epsilon_{\rm ff}$ leads to faster gas consumption and cloud dispersal due to earlier and stronger stellar feedback and thus shorter lifetime. In the Nofeed run, GMCs live substantially longer than in other models, with over ten surviving beyond $20 \, \mathrm{Myr}$. In the SN run, GMCs with lifetimes below $20 \, \mathrm{Myr}$ tend to last longer than in the fiducial run, whereas those exceeding $20 \, \mathrm{Myr}$ have shorter lifetimes. Conversely, in the Rad run, GMCs with lifetimes below $20 \, \mathrm{Myr}$ show distributions similar to the fiducial run, while those above $20 \, \mathrm{Myr}$ persist longer due to the absence of SN feedback.
    
    \item Unlike the simple exponential decay form reported in previous studies, we find that the cumulative distribution of GMC lifetimes is better described by a two-component exponential decay. This distinction arises from different galactic dynamics modes at different galactic location. In the inner regions, strong tidal disturbances prolong GMC lifetimes, keeping them marginally unbound for most of their evolution. In contrast, GMCs in the outer regions have shorter lifetimes, remaining gravitationally bound as they collapse due to weaker tidal forces.

    \item We capture the entire life cycles of GMCs in our simulated galaxies, from accretion and star formation to stellar feedback-driven cloud dispersal. We find that the accretion process continuously supplements the gas content throughout most of GMCs' life cycles, making the cloud mass higher and virial parameter lower in the early stage. When the cloud mass reaches its peak, feedback from young stars disperses the cloud, reduces the cloud mass and inflates the virial parameter. Interestingly, there is a positive correlation between GMC lifetimes and GMC maximum gaseous masses, which stands true regardless of star formation and stellar feedback models. 

    \item The GMC-scale SFE varies with GMC properties and galactic environments. Consistent with previous isolated GMC studies, a positive correlation is observed between baryonic surface densities and GMC-scale SFEs across all model variations. Distinct stellar feedback mechanisms play specific roles: early feedback inhibits the growth of dense clumps, while supernova feedback disperses extremely dense clumps. Galactic shear also regulates GMC-scale star formation, exerting an influence comparable to stellar feedback. When considering the combined effects of stellar feedback and galactic shear, the GMC-scale star formation efficiency can be described by $\varepsilon_{\mathrm{int}}=\left(1+(\Sigma_{\mathrm{fb}}+\Sigma_{\mathrm{shear}})/\Sigma_{\mathrm{max}}\right)^{-1},$ where $\Sigma_{\mathrm{fb}} \sim 2.8\times 10^3 \, M_{\odot} \, \mathrm{pc}^{-2}, \, \Sigma_{\mathrm{shear}} \sim 3.9 \times 10^3 \, M_{\odot} \, \mathrm{pc}^{-2}$. This suggests that external galactic environments play a similar, if not bigger, role than internal stellar feedback in shaping GMC life cycles.

    \item Analyzing GMCs with a fixed $10 \, \mathrm{Myr}$ lifetime across different feedback models reveals a consistent trend: $\varepsilon_{\mathrm{ff}}$ generally increases during the early stages and decreases in the later stages of GMC life cycles. More massive GMCs sustain higher $\varepsilon_{\mathrm{ff}}$ (above $2$) throughout their lifetimes, whereas less massive GMCs exhibit very low $\varepsilon_{\mathrm{ff}}$ (below $2$) in their early stages.

\end{itemize}

\begin{acknowledgements}
We thank Volker Springel for giving us access to \textsc{arepo}.
YN is grateful to Yunwei Deng, Aaron Smith, Josh Borrow, Greg Bryan, and Florent Renaud for useful discussions. HL is supported by the National Key R\&D Program of China No. 2023YFB3002502, the National Natural Science Foundation of China under No. 12373006, and the China Manned Space Program through its Space Application System. YN was a visiting student at the Massachusetts Institute of Technology sponsored by the Zheng Gang Scholarship for Overseas Study. 

\end{acknowledgements}

\bibliographystyle{aa}
\bibliography{aana}

\end{document}